\RequirePackage{fix-cm}
\documentclass{svjour3}                     
\smartqed  
\usepackage{graphicx}
\usepackage{amsmath}
\usepackage[T1]{fontenc}
\usepackage{amsmath}
\usepackage[utf8]{inputenc}
\usepackage{mathtools}
\usepackage{latexsym}
\DeclareUnicodeCharacter{2212}{-}
\begin{document}

\title{On the dynamics of gravity induced wave function reduction}


\author{Faramarz Rahmani\and Mehdi Golshani}
\institute{F. Rahmani \at
Department of Physics, School of Sciences, Ayatollah Boroujerdi University, Boroujerd, Iran\\
              School of Physics, Institute for Research in Fundamental Science(IPM), Tehran, Iran\\
              Tel.: +98-21-22180692,
              Fax: +98-21-22280415\\
              \email{faramarz.rahmani@abru.ac.ir; faramarzrahmani@ipm.ir}           
              \and 
               M. Golshani \at
              School of Physics, Institute for Research in Fundamental Science(IPM), Tehran, Iran\\
              Department of Physics, Sharif University of Technology, Tehran, Iran\\
              Tel.:+98-21-66022718, Fax.:+98-21-66022718\\
              \email{golshani@sharif.edu} }   
\date{Received: date / Accepted: date}

\maketitle

\begin{abstract}
In this study, we use the concept of Bohmian trajectories to present a dynamical and deterministic interpretation for the gravity induced wave function reduction. We shall classify all possible regimes for the motion of a particle, based on the behavior of trajectories in the ensemble and under the influence of quantum and gravitational forces. In the usual approaches all information are obtained from the wave function evolution. But, on the basis of Bohm's deterministic quantum theory, we can investigate the motion of particle during the reduction processes. This leads to analytical and numerical results for the reduction time and equation of motion of the particle. In this regard, a new meaning will be provided for the reduction time.
\end{abstract}

\section{Introduction}
\label{intro}
A fundamental question that can be asked is that what criterion can determine the boundary between the quantum and classical worlds?. In other words, how we can say an object is a macroscopic or microscopic object?. By macroscopic objects we mean those objects that obey the laws of classical mechanics while microscopic objects are governed by the rules of quantum mechanics. It seems that by increasing the mass of an objects its classical behavior increases. In other words, its position and momentum are precisely determined, and we do not say a particle is in several places instantaneously. In classical world, we do not see a macroscopic object in a quantum superposition. Interference patterns are not seen for a macroscopic object. What we receive from experiments and our daily experiences demonstrates that by going to molecular and atomic scales, quantum behavior increases. In other words, we observe interference patterns, uncertainty in the position and momentum and indeterminism. In gravity induced theories, the mass of the particle which is an objective property of the particle is the criterion for determining the boundary between the quantum and classical worlds. The fundamental affecting force which reduces the wave function is gravity. Thus such approaches are usually called "objective gravitational wave function reduction". What happens in objective gravitational reduction is different from the usual instantaneous collapse of the usual quantum mechanics. There, due to the linearity of the Schr\"{o}dinger equation, there is no explanation for the wave function collapse. Thus, the unjustified collapse of the wave function has been accepted as one of the postulates of standard quantum mechanics. Since, by reducing the wave function, quantum superposition breaks down, it is needed to modify the Schr\"{o}dinger equation by nonlinear terms like as gravitational potentials. Previous gravity induced studies of some authors, shows that the reduction processes takes place during a period of time and not instantaneously. Many studies have been done in this regard in the framework of standard or usual quantum mechanics.\cite{RefK1,RefK2,RefK3,RefD1,RefD2,RefD3,RefP1,RefP2,RefP3}.
We want to study gravity induced wave function reduction in the Bohmian context. Because, in Bohmian quantum mechanics, there is room for concepts like, trajectory, velocity, force, etc. Bohmian mechanics is a deterministic quantum theory which provides an objective realist view of the quantum world. The wave function of the system and the trajectories of the particle provide a complete description of the system. In fact, a system consists of the particle and wave function. The system can have simultaneous position and momentum but they can not be simultaneously measured.\cite{RefB,RefI,RefU,RefH,RefC,RefPet}. In standard quantum mechanics, we get all information from the wave function and there is no room for defining such dynamical variables before any measurement. In Bohmian quantum mechanics, indeterminism comes from ignorance of the particle's initial position (hidden variable) and is not an intrinsic aspect of the theory.\cite{RefH,RefM}. An overview of the hidden variable theories is available in the refs.\cite{RefBal,RefBell}. Therefore, we hope to provide a deterministic formalism for the wave function reduction.\par 
We introduce some of the most important works in standard quantum mechanics. Some of them are not in the context of gravity induced wave function reduction. But, they are outstanding works. For example the GRW model which is based on suddenly jumps of the wave function and provides a universal dynamics for the evolution of the system.\cite{RefGir1,RefGir2,RefBassi}.\par 
One of the important works in the field of gravity-induced wave function reduction was done by Karolyhazy based on gravitational concepts and spacetime physics.\cite{RefK1}. After that, another studies were done of which the two significant ones are the works of the Penrose and Diosi. 
The Penrose explanations are based on general principles of physics such as the general covariance principle and the equivalence principle.\cite{RefP1,RefP2}. In these approaches, there is an uncertainty in the structure of spacetime(an ensemble of metrics is considered) which causes the uncertainty in the position of the particle or object. Reducing the uncertainty of the space-time structure, leads to the reduction of the matter wave function. Of course, these approaches have differences in mathematical and physical details. But for a point particle their results are identical. In contrast to these approaches, we  shall study the wave function reduction in a single spacetime but with an ensemble of trajectories. To compare the results of our work with the results of previous works in the future, we review the main strategy of the first work of Diosi shortly.\cite{RefD1}.\par 
A Gaussian wave packet with the initial width $\sigma_0$, spreads over time (quantum mechanical behavior). In order to have stationary wave packet, the particle mass must be equal to a specific critical value to produce the required gravity to prevent the dispersion of the wave packet.
This can be achieved by a modified Schr\"{o}dinger equation which contains the self-gravitational potential of the particle. The new equation is called the Schr\"{o}dinger-Newton equation which for a single particle with distribution $\rho = \vert \psi(\mathbf{x},t) \vert ^2$ is:
\begin{equation}\label{sn}
i\hbar \frac{\partial\psi(\mathbf{x},t)}{\partial t}=\left(-\frac{\hbar^2}{2m}\nabla^2 -Gm^2 \int \frac{\vert \psi(\mathbf{x}^\prime,t)\vert^2}{\vert \mathbf{x}^\prime -\mathbf{x} \vert} d^3 x^\prime\right) \psi(\mathbf{x},t).
\end{equation}
Minimizing the Schr\"{o}dinger-Newton Hamiltonian functional equation for a stationary wave packet, gives a relation between the critical mass of the particle and the characteristic width of its associated stationary wave packet. In other words, the value of $\sigma_0$ for which we can have a stationary wave packet is determined by the value of the particle mass and is equal to:
\begin{equation}\label{Dio}
\sigma_{c} = \frac{\hbar^2}{Gm^3}
\end{equation}
An important point is that, the wave function and its width are not observable but relation (\ref{Dio}) expresses the characteristic  width of the stationary wave packet  in terms of particle mass which is an objective measurable property and constants of nature, $\hbar$ and $G$. This relation also gives the boundary between classical and quantum worlds. In other words, by using the relation (\ref{Dio}), we can define a critical mass for the transition from quantum domain to the classical domain for a point like particle. We have:
\begin{equation}\label{mc}
m_c=\left(\frac{\hbar^2}{G\sigma_c }\right)^{\frac{1}{3}}
\end{equation}
By this relation at hand, we can say that the particles with masses greater than the critical mass represents more macroscopic behavior and for the particles with masses less than the critical mass, micro behaviors increases. Thus, an ambiguous collapse hypothesis transforms to an objective issue. \par 
It is appropriate to have a clarification about the  self-gravity for a pointlike particle. The meaning of the self-gravity is explained on the basis of uncertainty in the position of the particle or object. We know that a particle is detected around the point $\mathbf{x}$ in the configuration space at the time $t$ with the probability density $\rho (\mathbf{x},t)=\vert \psi \vert ^2$. Here, the mass distribution of a point particle seems like an extended mass distribution in the configuration space of the particle. Hence, we conclude that the definition of self-gravity is possible for a point particle in a quantum mechanical sense. This argument is also  true for extended objects. \par
Usually, when we talk about the wave function reduction, our mind turns to the wave function reduction through a measurement process. In fact, reduction through a measurement can also be explained by considering the gravitational effects of the measuring apparatus which its wave function is entangled with the wave function of the quantum system. The self-gravity due to the apparatus reduces its own wave function and consequently the entangled wave function of the system. For details see \cite{RefP2}.\par 
Now, it is good to have a quick look at the essentials of Bohmian quantum mechanics. 
In Bohmian quantum mechanics, the wave function is considered as a pilot wave which guides a particle on an ensemble of trajectories in a deterministic manner. But, due our ignorance to the initial position which is considered as hidden variable in the Bohmian quantum mechanics, we do not know on which trajectory the particle is. The primary approach of the Bohm was the substitution of the polar form of the wave function, $\psi(\mathbf{x},t)=R(\mathbf{x},t)\exp(i\frac{S(\mathbf{x},t)}{\hbar})$, into the Schr\"{o}dinger equation which leads to a modified Hamilton-Jacobi equation as:
\begin{equation}\label{hamilton}
\frac{\partial S(\mathbf{x},t)}{\partial t}+\underbrace{\frac{(\nabla S)^2}{2m}+V(\mathbf{x})+Q(\mathbf{x})}_{E}=0
\end{equation}
The last term in the above equation is the non-relativistic quantum potential which is given by:
\begin{equation}\label{potential}
Q=-\frac{\hbar^2 \nabla^2 R(\mathbf{x},t)}{2mR(\mathbf{x},t)}=-\frac{\hbar^2 \nabla^2 \sqrt{\rho(\mathbf{x},t)}} {2m\sqrt{\rho(\mathbf{x},t)}}
\end{equation}
The quantum force exerted on the particle is defined as $\mathbf{f}=-\nabla Q$. The function $S(\mathbf{x},t)$, is the action of the particle which appears in the phase of the wave function. It propagates like a wave front in the configuration space of the particle. The momentum field of the particle is a vector field orthogonal to the $S(\mathbf{x},t)$ which is obtained by using the relation $\mathbf{p}=\nabla S(\mathbf{x},t)$.  
The position of the particle satisfies the guidance equation: 
\begin{equation}\label{guidance}
\frac{d\mathbf{x}(t)}{dt}=\left(\frac{\nabla S(\mathbf{x},t)}{m}\right)_{\mathbf{X}=\mathbf{x}(t)}.
\end{equation}
With the initial velocity $\mathbf{v}_0 = \frac{\nabla S_0}{m}=\frac{\mathbf{p}_0}{m}$, and the initial position $\mathbf{x}_0$, the system evolves in a deterministic manner according to the guidance equation (\ref{guidance}). But practically, we are faced with a distribution of the initial positions and velocities. Thus, there is always an uncertainty in the prediction of evolution of the system. In the presence of the external potential $V(\mathbf{x})$, the dynamics of the system is obtained through the relation 
\begin{equation}
\mathbf{f}= m\frac{d^2 \mathbf{x}}{dt^2}=-\nabla Q(\mathbf{x})-\nabla V(\mathbf{x}).
\end{equation}
The expression $\mathbf{X}=\mathbf{x}(t)$ in relation (\ref{guidance}) means that among all possible trajectories, one of them is chosen.(We do not know which trajectory is chosen). In following, we first discuss about the physical concepts of the issue in the Bohmian context. Then, we shall derive the  reduction criterion and the equations of motion in the following sections.
\section{Bohmian trajectories and objective criterion}
\label{sec:1}
First, let us have a look at the free particle motion in Bohmian quantum mechanics. Then, we shall argue how does gravity affect its evolution. The particle is guided by the wave packet and moves according to the guidance equation (\ref{guidance}). The amplitude of the Gaussian wave packet with zero initial group velocity is:
\begin{equation}\label{rf}
R= (2\pi \sigma^2)^{-\frac{3}{4}} e^{-\frac{\mathbf{x}^2}{4\sigma^2}}
\end{equation}
where, $\sigma$ is the random mean square width of the packet at  the time $t$. It is equal to:
\begin{equation}\label{width}
\sigma=\sigma_0 \sqrt{1+\frac{\hbar^2 t^2}{4m^2 \sigma_0^4}}
\end{equation}
and $\sigma_0$ is the width of the wave packet at $t=0$. If $m \rightarrow \infty$ in the relation (\ref{width}), or at the times $t\simeq 0$, the spreading of the wave packet is negligible. Then, for $t\simeq 0$, we have  $\sigma \approx \sigma_0$. But by increasing $t$ in the numerator of $\frac{\hbar^2 t^2}{4m^2 \sigma_0^4}$ in the relation (\ref{width}), dispersion will be significant again. So, we need a balancing term in the particle dynamics. In the gravitational approaches, the self-gravity (gravity due to the distribution $\rho=\vert \psi \vert ^2$), prevents a more dispersion of the wave packet and for a specific mass we have equilibrium of the quantum and gravitational forces. The criterion (\ref{Dio}) is obtained through the balancing of these forces in the language of standard quantum mechanics. We shall derive it through the concepts of Bohm's deterministic quantum theory.\par
\begin{figure}[ht] 
\centerline{\includegraphics[width=6cm]{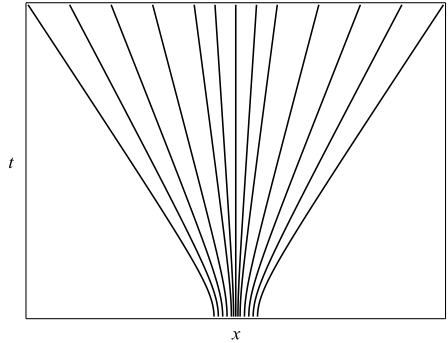}}
\caption{An ensemble of divergent trajectories is produced by the quantum force. \label{fig:1}}
\end{figure}

 
It has been demonstrated in ref \cite{RefH} that the trajectory of a free particle which is guided by a Gaussian wave packet is:
\begin{equation}\label{tra}
\mathbf{x}(t)=\mathbf{x}_{0}\left(1+(\frac{\hbar t}{2m\sigma_0^2})^2\right)^{\frac{1}{2}}
\end{equation}
where, $\mathbf{x}_{0}$ denotes the initial position of the particle. In figure (\ref{fig:1}), we have plotted trajectories in the $(t,x)$ plane. Due to uncertainty in the position of the particle, we are faced with an ensemble of trajectories.  
\par  
Bohmian quantum potential for this system is:
\begin{equation}\label{qf}
Q=-\frac{\hbar^2 \nabla^2R}{2mR}= \frac{\hbar^2}{4m\sigma^2}\left(3-\frac{\mathbf{x}^2}{2\sigma^2} \right)
\end{equation}
Hence, Bohmian quantum force is:
\begin{equation}\label{ff}
\mathbf{f}=-\nabla Q = \frac{\hbar^2}{4m\sigma^4}\mathbf{x}
\end{equation}
It is clear that the quantum force is repulsive. The figure (\ref{fig:1}), shows that for initial positions with $\mathbf{x}_{0}\neq0$, the trajectories are not straight lines and the particle is affected by a quantum force. This leads to more separation between trajectories over time. 
From  the usual point of view, the wave packet spreads. It is usually justified by using the Heisenberg uncertainty principle which leads to the statement that the probability of finding particle around the initial position decreases with time. Its correspondence in Bohmian quantum mechanics, is that the repulsive quantum force derives the particle a way from the center of the wave packet. Therefore the deviation between  associated trajectories increases. 
For returning curved trajectories to the straight parallel trajectories, we need an agent to keep the deviation between trajectories constant. It seems that the self-gravity of the particle is able to do this. In other words, gravity prevents more dispersion and consequently more uncertainty in the position of the particle. By increasing the mass of the particle, its self-gravity increases. Then, the quantum force and gravitational force can be equal for a specific value of mass and the deviation between trajectories remains constant. In this situation, we are on the threshold of the classical world. We can call it transition regime. We expect that for the particles with masses greater than  the specific critical mass $m_c$, self-gravitation increases and trajectories become convergent. Then, we are moving more and more towards the classical world i.e, more certainty and the wave function reduces to a delta Dirac function which we shall demonstrate it. Also, for $m<m_c$ the quantum features increases and  the uncertainty in the particle position increases. \par 
Fortunately, it can be proved that Bohmian trajectories do not cross each other.\cite{RefH}. Thus,
these trajectories can be considered as a congruence in space and we can define a deviation vector between trajectories without any trouble.
In fact, the single-valuedness of the wave function for a closed loop leads to:
\begin{equation}
\Delta S = \oint_c dS = \oint_c \nabla S \cdot d\mathbf{x}= \oint_c \mathbf{p}\cdot d\mathbf{x}=nh
\end{equation}
Then, the net phase difference around a closed loop is constant and for every instant at the position $\mathbf{x}$, the momentum $\mathbf{p}=\nabla S$ is single-valued. 
 \par
\begin{figure}[ht] 
\centerline{\includegraphics[width=6cm]{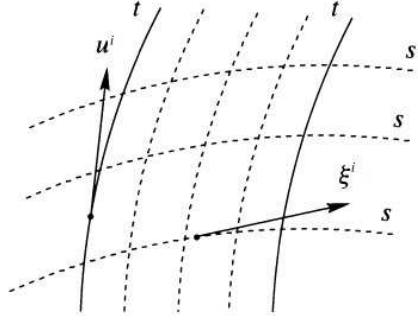}}
\caption{An ensemble of trajectories and the deviation vector $\xi^i$ between two nearby trajectories are shown.  \label{fig:2}}
\end{figure}

 
Now, consider an ensemble of non crossing trajectories of a particle. We represent trajectories  as $x^i(t,s)$, where $t$ is time and $s$ labels the trajectories. See figure (\ref{fig:2}).
The parameters $t$ and $s$ are independent parameters. The vector $u^i$ in the figure (\ref{fig:2}), denotes the velocity of the particle on the $i$th trajectory in the ensemble. The deviation vector between the two nearby trajectories is:
\begin{equation}\label{dev}
\xi^i =\frac{\partial x^i(t,s)}{\partial s}
\end{equation}
which is tangent to the curves parameterized by $s$.
The relative acceleration between two nearby trajectories becomes:
\begin{equation}\label{adev}
\ddot{\xi^i}=\frac{\partial^2 \xi^i}{\partial t^2}
= \frac{\partial ^2}{\partial t^2}\left(\frac{\partial x^i(t,s)}{\partial s}\right)=\frac{\partial}{\partial s}\left( \frac{\partial^2 x^i(t,s)}{\partial t^2}\right)=\frac{\partial a^i}{\partial s}
\end{equation}
Where, $a^i$ is the total acceleration of the particle when it is on the $i$th trajectory in the ensemble. Here, this acceleration is due to the self-gravitational and quantum forces. The quantum force is $-\nabla Q$ where the quantum potential $Q$ has the dimension of energy and the acceleration associated with it is $-\frac{\nabla Q}{m}$. Therefore, we have:
\begin{equation}\nonumber
\frac{\partial^2 \xi^i}{\partial t^2}=-\frac{\partial}{\partial s}\left(\frac{1}{m}\frac{\partial U}{\partial x^i} + \frac{1}{m}\frac{\partial Q}{\partial x^i}\right)
\end{equation}
where, by using the $\frac{\partial}{\partial s} = \xi^j \frac{\partial}{\partial x^j}$, along the curves parametrized by $s$, takes the form:
\begin{equation}\label{eq1}
\frac{\partial^2 \xi^i}{\partial t^2}=-\xi^j \frac{\partial}{\partial x^j} \left(\frac{1}{m}\frac{\partial Q}{\partial x^i}+\frac{\partial \phi}{\partial x^i}\right)
\end{equation}
We conclude from this equation that for having zero deviation acceleration or parallel trajectories,
one of the possibilities is
 \begin{equation}\label{eq2}
\frac{\partial \phi}{\partial x^i}=-m \frac{\partial Q}{\partial x^i}, \qquad i=1,2,3
\end{equation}
which imply the equality of the self-gravitational force and the Bohmian quantum force.\\
In a three dimensional notation it takes the form
\begin{equation}\label{eq3}
\nabla Q=-m\nabla \phi
\end{equation} 
By imposing ($\nabla \cdot $) on the both side of above equation, we have: 
\begin{equation}\label{eq4}
\nabla^2 Q =- m\nabla^2 \phi 
\end{equation}
where, by using the Poisson equation
\begin{equation}\nonumber
\nabla ^2 \phi =4\pi Gm\rho ,
\end{equation}
it becomes:
\begin{equation}\label{eq44}
\nabla^2 Q = 4\pi Gm^2 \rho
\end{equation}
It leads to:
\begin{equation}\label{eq5}
\frac{\hbar^2}{2m^2}\nabla^2\left(\frac{\nabla^2 \sqrt{\rho}}{\sqrt{\rho}}\right)=4\pi G m\rho
\end{equation}
Relations (\ref{eq2}), (\ref{eq3}), (\ref{eq4}) and (\ref{eq5}) are the mathematical expressions of our previous physical arguments. We see that how  the transition from quantum domain to the classical world is represented by a dynamical equilibrium between the quantum and self-gravitational forces.
Relation (\ref{eq5}) is a nonlinear differential equation for the transition from quantum domain to the classical domain. \par 
Now, we want to obtain the famous criterion (\ref{Dio}) in the Bohmian context. We can do it in different ways. For example, we determine the average quantum potential and the self-gravitational potential as the functions of  the width $\sigma_0$. Then, we find the characteristic width for which gravity and quantum are balanced. Suppose that due to the equality of the self-gravitational force and the quantum force, our Gaussian wave packet  with the initial width $\sigma_0$, is non-dispersive. It is appropriate to work in spherical coordinate for simplifying calculations. 
If we obtain the quantum potential for a stationary  Gaussian wave packet with the amplitude $R(r)= (2\pi \sigma_0^2)^{-\frac{3}{4}}e^{-\frac{r^2}{4\sigma_0^2}}$, where, $\rho(r)= R^2(r)=(2\pi )^{-\frac{3}{2}} \sigma_0^{-3} e^{-\frac{r^2}{2\sigma_0^2}} $, we get:
\begin{equation}\label{sq}
Q^{(s)}=-\frac{\hbar^2}{2m}\frac{\nabla^2 R}{R}= -\frac{\hbar^2}{2m}\frac{1}{r}\frac{\partial}{\partial r}\left(r\frac{\partial R}{\partial r}\right)=\frac{\hbar^2}{8m\sigma_0^4}\left(6\sigma_0^2-r^2\right)
\end{equation}
The average quantum potential becomes:
\begin{eqnarray}\label{aq}
\langle Q^{(s)}\rangle &=&\int_{0}^{\infty} \rho Q^{(s)} 4\pi r^2 dr = 4\pi\int_{0}^{\infty} R^2  \left(-\frac{\hbar^2}{2m}\frac{\nabla^2 R}{R}\right) r^2 dr  \nonumber \\ &=&   \lim_{r\rightarrow \infty} \left(\frac{3\hbar^2}{8m\sigma_0^2} erf \left(\frac{\sqrt{2}r}{2\sigma_0}\right) \right)\approx \frac{\hbar^2}{m\sigma_0^2}
\end{eqnarray}
where, we have used the relation (\ref{sq}) to get this result. \footnote{We have used $\lim_{r\rightarrow \infty } erf \left(\frac{\sqrt{2}r}{2\sigma_0}\right) =1$ in the relation (\ref{aq}).} The gravitational self energy for a particle with mass $m$ is defined as:
\begin{equation}\label{self}
U_{g}(\mathbf{x})=-Gm^2 \int \frac{\vert \psi(\mathbf{x}^\prime,t)\vert^2}{\vert \mathbf{x}^\prime -\mathbf{x} \vert} d^3 x^\prime
\end{equation}
where, $\vert \psi(\mathbf{x}^\prime,t)\vert^2 = R^2(\mathbf{x}^\prime)$. Just a few calculation in spherical coordinates for a particle with a stationary Gaussian wave packet, gives:
\begin{equation}\label{pot2}
U_{g}^{(s)}(r)=- \int_{0}^{r} \frac{Gm^2}{r^{\prime}}\rho(r^{\prime}) 4\pi r^{\prime^2} dr^{\prime}=\sqrt{\frac{2}{\pi}}\frac{Gm^2}{\sigma_0}\left(1-e^{-\frac{r^2}{2\sigma_0^2}}\right)
\end{equation}
The average gravitational self energy of the Gaussian distribution, with the width $\sigma_0$, is equal to: 
\begin{equation}\label{ag}
\langle U _{g}^{(s)}\rangle = \int_{0}^{\infty} \rho(r) U_{g}^{(s)}(r) 4\pi r^2 dr \approx -\frac{Gm^2}{\sigma_0}
\end{equation}
In Bohmian quantum mechanics for a stationary state with one degree of freedom, the momentum of the particle is zero and the Hamilton-Jacobi equation does not include the kinetic term. \cite{RefH}. Thus, for our system, the average of the modified Hamilton-Jacobi equation becomes:
\begin{equation}\label{ah}
\langle E \rangle =\langle U _{g}^{(s)}\rangle + \langle Q^{(s)}\rangle
\end{equation}
Now, by substituting relations (\ref{aq}) and (\ref{ag}), into the relation (\ref{ah}) and minimizing the average energy of the system with respect to the $\sigma_0$, i.e,
\begin{equation}
\frac{\delta \langle E \rangle}{\delta \sigma_0}=\frac{\delta \langle Q^{(s)}\rangle}{\delta \sigma_0}+\frac{\delta \langle U _{g}^{(s)}\rangle}{\delta \sigma_0}=0,
\end{equation}
we obtain the characteristic width  for which the average gravitational force and quantum force are equal. In other words, we have:
\begin{equation}\label{dio}
 (\sigma_0)_{\text{critical}}= \frac{\hbar^2}{Gm^3}
\end{equation}
 \par 
We can also calculate the average quantum force and gravitational force directly. 
The quantum force becomes:
\begin{equation}\label{sf}
f_q^{(s)} =-\frac{\partial Q^{(s)}}{\partial r}=\frac{\hbar^2}{4m\sigma_0^4} r
\end{equation}
Now, the average quantum force in the ensemble is
\begin{equation}\label{asq}
\langle f_q^{(s)} \rangle = \int_{0}^{\infty} \rho(r) f_q^{(s)} 4\pi r^2 dr=\frac{1}{2}\sqrt{\frac{2}{\pi}} \frac{\hbar^2}{m\sigma_0^3} \approx \frac{\hbar^2}{m\sigma_0^3}
\end{equation}
In the same way, the self-gravitational force and its average are as bellow.
\begin{equation}\label{sg}
f_g^{(s)}=-\frac{\partial U^{(s)}(r)}{\partial r}=-\sqrt{\frac{2}{\pi}}\frac{Gm^2}{\sigma_0^3}r e^{-\frac{r^2}{2\sigma_0^2}}
\end{equation}
Then,
\begin{equation}\label{asf}
\langle f_g^{(s)} \rangle = \int_{0}^{\infty} \rho(r) f_g^{(s)} 4\pi r^2 dr=-\frac{1}{\pi}\frac{Gm^2}{\sigma_0^2} \approx -\frac{Gm^2}{\sigma_0^2}
\end{equation}
Equating the results of (\ref{asq}) and (\ref{asf}) leads to the criterion (\ref{dio}). 
We see that how the concepts of Bohmian quantum mechanics provide an imaginable and dynamical framework for studying the reduction processes. By the relation (\ref{dio}), we are at threshold of the classical world. 
Let's take a closer look at the behavior of trajectories in the following.
\section{The classifications of  particle dynamics}
\label{sec:2}
In the following, we classify all possible situations according to the evolution of the deviation vector. 
But, before we go any further, we rewrite the equation (\ref{eq1}) in a new form. 
First we define a quantity as:
\begin{equation}\label{xo}
\Omega_{ij}(x)=-\frac{\partial^2}{\partial x^j\partial x^i}\left(\frac{Q}{m}+\phi \right) \quad i=1,2,3
\end{equation}
Then, the equation (\ref{eq1}), takes the form:
\begin{equation}\label{xo2}
\ddot{\xi}^i=\Omega_{ij} \xi^j , \quad  i=1,2,3
\end{equation}
It is clear from the relation (\ref{xo}), that $\Omega_{ij}$ is symmetric i.e, $\Omega_{ij}=\Omega_{ji}$.
In the matrix representation this equation takes the form:
\begin{equation}\label{mat1}
\begin{pmatrix}
\ddot{\xi}^1 \\
\ddot{\xi}^2 \\
\ddot{\xi}^3  \\
\end{pmatrix}  = 
\begin{pmatrix}
\Omega_{11} & \Omega_{12} & \Omega_{13} \\
\Omega_{21} & \Omega_{22} & \Omega_{23} \\
\Omega_{31} & \Omega_{32} & \Omega_{33} \\
\end{pmatrix}
\begin{pmatrix}
\xi^1 \\
\xi^2 \\
\xi^3  \\
\end{pmatrix}
\end{equation}
It seems from matrix representation that we are in trouble for existing non-diagonal elements ($\Omega_{ij}, i \neq j$ ) and our arguments for reaching the relation (\ref{eq4}) is for an special case i.e, for which the matter distribution is isotropic and only diagonal elements are present. Fortunately, there is no need to worry. Because, in the study of wave function reduction we deal with a one particle system or a rigid body. Hence, thermodynamics properties like the pressure and other fluid properties like the viscosity and heat flows are meaningless, and non-diagonal elements are vanished. However, matrix representation  may be useful for further developments for example to study the behavior of a quantum fluid at the boundary of the classical world at least from the theoretical aspect of view. Perhaps, such generalization by matrix language may be more useful to the study of a quantum fluid evolution in an external gravitational field.
With these considerations, the relation (\ref{mat1}) reduces to:
 \begin{equation}\label{mat2}
\begin{pmatrix}
\ddot{\xi}^1 \\
\ddot{\xi}^2 \\
\ddot{\xi}^3  \\
\end{pmatrix}  = 
\begin{pmatrix}
\Omega_{11} & 0 & 0 \\
0 & \Omega_{22} & 0 \\
0 & 0 & \Omega_{33} \\
\end{pmatrix}
\begin{pmatrix}
\xi^1 \\
\xi^2 \\
\xi^3  \\
\end{pmatrix}
\end{equation}
where, $\Omega_{11}=\Omega_{22}=\Omega_{33}=\Omega$. For example in the Cartesian coordinates, for an isotropic distribution we have:
\begin{eqnarray}
\Omega_{11}=\Omega_{x}=-\left(\frac{1}{m}\frac{d^2 Q}{dx^2}+\frac{d^2 \phi(x)}{dx^2}\right)= -\left(\frac{1}{m}\frac{d^2 Q}{dx^2}+4\pi G m\rho(x)\right)\\
\Omega_{22}=\Omega_{y}=-\left(\frac{1}{m}\frac{d^2 Q}{dy^2}+\frac{d^2 \phi(y)}{dy^2}\right)= -\left(\frac{1}{m}\frac{d^2 Q}{dy^2}+4\pi G m\rho(y)\right)\\
\Omega_{33}=\Omega_{z}=-\left(\frac{1}{m}\frac{d^2 Q}{dz^2}+\frac{d^2 \phi(z)}{dz^2}\right)= -\left(\frac{1}{m}\frac{d^2 Q}{dz^2}+4\pi G m\rho(z)\right)
\end{eqnarray}
If, $\Omega_{x}=\Omega_{y}=\Omega_{z}=0$, then $\ddot{\xi}^i=0$, and we have parallel trajectories. The second derivative of quantum potential (relations(\ref{qf}) or (\ref{sq})) with respect to the position is always  negative. On the other hand, due to the positive nature of $\rho$, it is possible to have vanishing $\Omega=\left(\frac{1}{m}Q^{\prime\prime}+4\pi G m\rho\right)$ for a specific mass $m_c$ in the above relations. It may be asked that the distribution $\rho$ in this parenthesis is an exponential function which can be expanded as a power series of position. How can it be equal with the second derivative of the quantum potential $Q$ which is just the second order function of position? We can answer to this question in two ways. In fact,we know the maximum probability of finding particle is in the interval $-\sigma_0 <\mathbf{x} <\sigma_0$. If the particle is at $\mathbf{x}=0$ or $\mathbf{x}=\pm \sigma_0$, the factor $e^{-\frac{\mathbf{x}^2}{2\sigma^2}}$ takes the values between $e^{0}=1$ and $e^{-\frac{1}{2}}=0.6$. This means that the effect of this factor is a finite number multiplied by the term $\sigma_0^{-3}$ of the distribution $\rho =(2\pi )^{-\frac{3}{2}} \sigma_0^{-3} e^{-\frac{\mathbf{x}^2}{2\sigma_0^2}}$. On the other hand the second derivative of the quantum potential is $Q^{\prime \prime} \sim \frac{\hbar^2}{m \sigma_0^4}$. Now, the relation $\Omega=\left(\frac{1}{m}Q^{\prime\prime}+4\pi G m\rho\right)\approx (-\frac{\hbar^2}{m^2 \sigma_0^4} + \frac{Gm}{\sigma_0^3})=0$, also gives the famous criterion (\ref{Dio}) or (\ref{dio}). Another justification is also possible. In fact, the quantum potential $Q=\frac{\nabla^2 \sqrt{\rho}}{\sqrt{\rho}}$ is derived from the Bohm's primary approach. It has been demonstrated in refs \cite{RefShoja,RefAtigh,RefMach} that the Bohmian quantum potential contains infinite even order terms of Laplacian operator and can be written in an exponential form $Q \propto e^{\mathcal{Q}}$. In other words, the quantum potential is $Q=Q^{(0)}+Q^{(2)}+Q^{(4)}+\cdots = Const +\frac{\nabla^2 \sqrt{\rho}}{\sqrt{\rho}}+\frac{\nabla^4 \sqrt{\rho}}{\sqrt{\rho}}+\cdots  $. The original potential of Bohm is $Q^{(2)}$ and $Q^{(0)}$ is a constant answer. Here, $\nabla^4$ is a notation for $\nabla \cdot \nabla (\nabla \cdot \nabla)$. Only even order derivatives are appeared in the extended Bohmian quantum potential. Naturally, these higher order terms can be compared with the higher order terms of the distribution $\rho$ which are even order terms of position. It is  not necessary to discuss this issue any further. Because the approximation $\rho \approx \frac{1}{\sigma_0^3}$, leads us to the desired result. \par
Based on the behavior of the $\Omega_{ij}$, we have three regimes for which the acceleration of the deviation vector can be zero($\ddot{\xi}^i=0$), positive($\ddot{\xi}^i>0 $) and negative ($\ddot{\xi}^i<0$). According to these discussions and arguments, we conclude that the deviation equation (\ref{eq1}) or (\ref{xo2}) can be written in the form:
\begin{equation}\label{acom}
\ddot{\xi}^i = -\left(\frac{\nabla^2 Q}{3m}+\frac{4\pi Gm }{3}\rho \right)\xi^i= -\frac{4\pi Gm}{3}\left(\frac{\nabla^2 Q}{4\pi Gm^2}+\rho \right)\xi^i, \quad i=1,2,3
\end{equation}
Now, we want to argue that for given $\sigma_0$, at $t=0$, how will the system evolves for different values of mass?
\subsection{\textbf{Transition regime}}
Consider a Gaussian wave packet with the initial width $\sigma_0$. In the absence of gravity we expect that the wave packet to disperse. But, if the mass of the particle is so large that the needed self-gravitational force for the balancing of the quantum force can be provided, i.e, $m=m_c$, then the  wave packet does not disperse. \par 
In this situation, the quantum force and the self-gravitational force are equal and this leads to $\Omega_{ii}=0$ for any direction. Then, the deviation equation becomes:
\begin{equation}
\ddot{\xi}^i=0, \qquad i=1,2,3 
\end{equation}
with the solutions
\begin{equation}
\xi^i=\dot{\xi}^i(0)t +\xi^i(0)
\end{equation}
If, we have $\dot{\xi}^i(0)\neq 0$, the trajectories also remain parallel. Because, the relative deviation velocity vector remains constant and trajectories do not diverge. The figure (\ref{fig:1}) and the equation (\ref{tra}) show that at $t\approx 0$ the trajectories are parallel. According to our arguments this parallelism is maintained by the equality of quantum force and self-gravitational force for a critical mass. In the usual language, we say that the wave packet is stationary or non-dispersive.
We may have $\dot{\xi}^i(0)= 0$. In this case, the trajectories are parallel and relative separation between them remains constant and the relative deviation velocity becomes zero. 
\subsection{\textbf{Quantum-Dominant regime} ($\ddot{\xi}^i>0 $)}
Now, consider a wave packet with the initial width $\sigma_0$. According to the relations (\ref{sf}) and (\ref{sg}), the quantum force is proportional to $1/m$ and the gravitational force is proportional to $m^2$. Thus, we expect that by decreasing the mass of the particle with respect to the critical mass, quantum force overcomes the self-gravitational force. Then, quantum effects become significant and the particle gains quantum mechanical dynamics rather than the classical one. In other words, a particle with mass less than the critical mass associated to the given $\sigma_0$, can not produce enough gravity to prevent wave packet dispersion. Thus, the uncertainty in the position of the particle increases or the wave packet spreads. If the mass is so tiny relative to the critical mass, then we can ignore the gravitational force practically. Such ideal regime can be called quantum-dominant regime. In this regime, the separation between trajectories increases ($\ddot{\xi}^i>0 $), and the deviation vector obeys the equation:
\begin{equation}\label{EQD}
\ddot{\xi}^i = - \frac{\nabla^2 Q}{3m} \xi^i  \qquad i=1,2,3 
\end{equation}
Here, there is nothing to prevent more deviation between trajectories. 
Since, for the Gaussian wave packet we have always $\nabla^2 Q <0$, then we conclude that, $\ddot{\xi}^i>0 $ in this regime.\par 
By using the dynamical equation $f_q=m\ddot{r}$, the equation of motion in the spherical coordinates becomes
\begin{equation}\label{rep}
\ddot{r}=\frac{\hbar^2}{4m^2\sigma^4(t)}r
\end{equation}
\begin{figure}[ht] 
\centerline{\includegraphics[width=7cm]{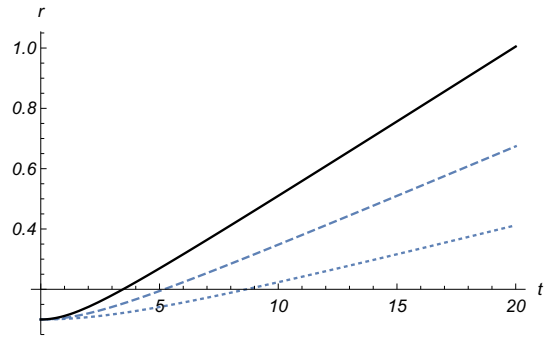}}
\caption{In the absence of gravity(quantum-dominant regime), the particle moves to infinity due to the repulsive nature of the quantum force. Here, we have worked in the Plank unit $\hbar=G=c=1$. \label{fig:3}}
\end{figure}
To plot this diagrams, we considered a fixed $\sigma_0$ and different masses with zero initial velocity. By decreasing the particle mass, its time-space diagram approaches the vertical axes. The repulsive nature of the quantum force has been shown in this figure.
\subsection{\textbf{Gravity-Dominant regime}}
As we said before, mass appears in the denominator of the quantum force while for the gravitational part it appears in the numerator. Thus, by increasing the mass of the particle, quantum force tends to zero. Now, consider the situation in which the  mass of the particle is so large relative to the critical mass that we can ignore the quantum force practically. 
As an extreme limit, when the mass of the particle is infinitely large, then its associated $\sigma_0$ tends to zero according to the famous criterion. Consequently, we shall have 
\begin{equation}\label{hj}
\rho =\lim_{\sigma_0 \to 0} (2\pi )^{-\frac{3}{2}} \frac{1}{\sigma_0^{3}} e^{-\frac{\mathbf{x}^2}{2\sigma_0^2}}  \longrightarrow \delta (\mathbf{x})  
\end{equation}
In this regime, the deviation equation (\ref{acom}) becomes:
\begin{equation}\label{neg}
\ddot{\xi}^i = -\frac{4\pi G m\rho(\mathbf{x})}{3}\xi^i , \qquad i=1,2,3
\end{equation}
If, $\sigma_0 \rightarrow 0$ ($m \rightarrow \infty$), then according to the (\ref{hj}), $\rho \rightarrow \delta(\mathbf{x})$. Thus, we will have:
\begin{equation}
\ddot{\xi}^i =-\frac{4\pi G m\delta(\mathbf{x})}{3} \xi^i , \qquad i=1,2,3
\end{equation}
which admits a classical singularity in the theory. In other words, trajectories lead to a singularity. However, infinite amount of mass is meaningless, but, the Dirac delta function is a appropriate limit practically. 
Due to the positive nature of $\rho$ in the relation (\ref{neg}), we conclude that always we have $\ddot{\xi}^i<0 ,(i=1,2,3)$ and the separation between trajectories decreases.\par 
That we see classical universe in a single state instead of being in a superposition, is due to the dominance of gravitational force over the quantum force not due to the spontaneous decay to a single state. In the Bohmian framework, there is no sharp distinction between the classical and quantum worlds. A quantum-dominant regime is that regime for which quantum force overcomes gravitational force. This does not mean that the gravitational force vanishes identically. In the same way, the gravity-dominant regime is that for which gravity overcomes the quantum effects. This does not mean that quantum force is exactly zero; rather it tends to zero. There is a dynamical equilibrium between this two forces for a critical mass. A little deviation from the critical mass, derives the system to one of these domains. To put it more clearly, consider a Gaussian wave packet with the initial width $\sigma_0$, if the particle that is going to be guided by this wave packet has the mass equal to the critical mass, the quantum and self-gravitational forces will be equal and we have a stationary system. For masses greater than the critical mass, classical behavior appears and for masses less than the critical mass, quantum features increase. Suppose that the mass of the particle is greater than the critical mass. We expect in this situation the particle falls toward the center of the distribution. In following, we study the dynamics of the falling particle in the gravity-dominant regime in which the quantum force has been ignored completely. However, when we ignore the quantum force, all particles fall to the center of distribution finally.
\section{The reduction time}
Consider the gravity-dominant regime. Our interpretation of the reduction time is as follow. In the absence of gravity, a particle takes some distances from the center of the wave packet due to the repulsive nature of the quantum force. On the other hand, the self-gravity defined based on this distribution, would localize the position of the particle. Here, by the reduction time, we mean the time   the particle falls from the around of center of the wave packet to its center. We can imagine  a Gaussian isotropic sphere with the radius $\sigma=\sigma(t)$. Gravity localizes the particle around the center of distribution. Thus, we expect massive particles are closer to the center of distribution and their associated wave packets are narrower. Thus, for the theoretical limit $m\rightarrow \infty$, the wave packet tends to the Dirac delta function in position space and the position of the particle becomes certain. We shall show that this happens through a dynamical equation of motion for the particle not through an ambiguous cut-off between the quantum and classical worlds. 
However, in following we have considered only the gravitational force, but we should not think that the quantum effects  have not been considered. Because, the Gaussian distribution in which the particle moves is due to the quantum effects. Now, let us study the dynamics of a falling particle in a Gaussian distribution.
We do it by writing the equation of motion of a particle, influenced by its own gravity, in spherical coordinates.   
The relation (\ref{sg}), represents the gravitational force for a stationary distribution with $\sigma=\sigma_0$. Thus, the general equation of motion of the particle with a varying $\sigma=\sigma(t)$, becomes
\begin{equation}\label{b1}
f_g=-\sqrt{\frac{2}{\pi}}\frac{Gm^2}{\sigma^3(t)}r e^{-\frac{r^2}{2\sigma^2(t)}}=m\ddot{r}
\end{equation}
Then, we have:
\begin{equation}\label{bb1}
\ddot{r}=-\sqrt{\frac{2}{\pi}}\frac{Gm}{\sigma^3(t)}r e^{-\frac{r^2}{2\sigma^2(t)}}
\end{equation}
The results of the numeric solutions of this equation have been plotted in the following figure for several masses. 
\begin{figure}[ht] 
\centerline{\includegraphics[width=8cm]{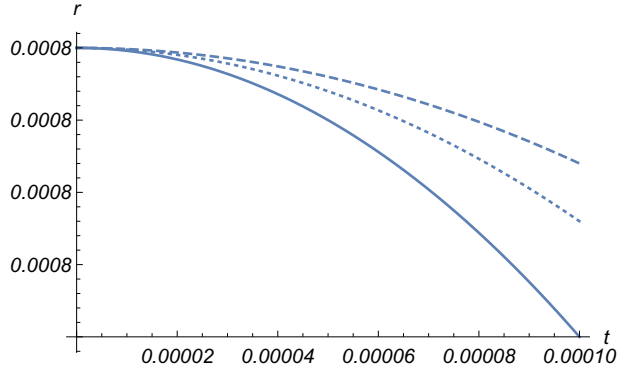}}
\caption{By increasing the mass of the particle for a fixed $\sigma_0$ and $\dot{r}(0)=0$ the reduction time i.e, the time that the particle falls to the center of the distribution, reduces. Here, we have worked in the Plank unit, $G=\hbar=c=1$.  \label{fig:4}}
\end{figure}
Here, we have considered the case of a fixed $\sigma_0$ and we have solved the equation (\ref{bb1}) for the several masses. We have assumed that the initial velocity of the particle is zero. We have worked in the Plank units ($G=\hbar=c=1$). Then for simplicity in plotting diagrams we considered the value of the critical mass equal to unity. This gives the characteristic width $\sigma_0=1$. Then, we solved the equation (\ref{bb1}) for the masses $2m_c, 4m_c$ and $5m_c$. The figure (\ref{fig:4}) shows that by increasing the particle mass, the falling time decreases. Thus, this is not a free falling motion and the falling time depends on the particle mass. This interpretation may seem a little strange. Because every particle eventually reaches the origin. In fact, for the light particles the reduction time is so long that we can not observe the wave function reduction practically. 
\subsection{Analytical results} 
Fortunately, the numeric solutions was obtained without any trouble. But, for having analytical solutions
 we need some assumptions. We shall see that the falling time can be interpreted as the reduction time and is consistent with the previous results of some authors. We assume that during the falling of the particle, $\sigma \approx \sigma_0$. This can be achieved by this condition that in the $\sigma(t)=\sigma_0 \sqrt{1+\frac{\hbar^2 t^2}{4m^2 \sigma_0^4}}$, we impose the condition $\frac{\hbar^2 t^2}{4m^2 \sigma_0^4} \ll 1$.
Thus, the equation of motion becomes:
\begin{equation}\label{b2}
\ddot{r}=\frac{du}{dr}u=-\sqrt{\frac{2}{\pi}}\frac{Gm}{\sigma^3_0} r e^{-\frac{r^2}{2\sigma^2_0}} \quad \text{for}\quad \sigma \approx \sigma_0
\end{equation}
which $u$ denotes the velocity of the particle when is on the one of the trajectories of ensemble.
 Now, we solve the equation (\ref{b2}) for having an estimation of reduction time. By integrating from the relation (\ref{b2}), we have:
\begin{equation}
\int_{u^{\prime}_0=0}^{u^{\prime}=u} u^{\prime}du^{\prime}=-\int_{r^{\prime}=r_0}^{r^{\prime}=r} \sqrt{\frac{2}{\pi}}\frac{Gm}{\sigma^3_0} r^{\prime} e^{-\frac{r^{\prime^2}}{2\sigma^2_0}} 
\end{equation}
or
\begin{equation}\label{vvv}
u^2 = 2\sqrt{\frac{2}{\pi}}\left(\frac{Gm}{\sigma^3_0}\right)\sigma_0^2 \left(e^{\frac{-r^2}{2\sigma_0^2}}-e^{\frac{-r_{0}^2}{2\sigma_0^2}}\right)
\end{equation}
Then,
\begin{equation}\label{vel}
u=\frac{dr}{dt}=\pm \sqrt{2}\left(\frac{2}{\pi}\right)^{\frac{1}{4}} \left(\frac{Gm}{\sigma^3_0}\right)^{\frac{1}{2}}\sigma_0 e^{-\frac{r_0^2}{\sigma_0^2}}\left(e^{\frac{r_0^2-r^2}{2\sigma_0^2}}-1 \right)^{\frac{1}{2}}
\end{equation}
The dimension of $\left(\frac{Gm}{\sigma^3_0}\right)^{\frac{1}{2}}$ is inverse of time and the dimension of $\sigma_0$ is length. Thus, the dimension of the right hand side of (\ref{vel}) is the dimension of velocity. This is the velocity of the particle as a function of position. Now, due to our ignorance about the initial conditions or uncertainty in the initial position $r_0$,  we are always faced with an ensemble of trajectories. In other words, the reduction equation (\ref{b2}) is deterministic intrinsically but we do not know initial data exactly. \par 
We know from the quantum mechanics that the maximum probability of finding the particle occurs within the width of the wave packet. 
In the presence of gravity we expect the particle to be localized around the center of wave packet further. On the other hand, the expansion $(e^{-\frac{r^2}{2\sigma_0^2}} \approx 1-\frac{r^2}{2\sigma_0^2}+\frac{r^4}{4\sigma_0^4}+\cdots)$  is convergent for the $r <\sqrt{2}\sigma_0$.
Thus, we expect that the interval $0 < r<\sigma_0$, is appropriate for investigation the problem.
Now, from the equations (\ref{vvv}) or (\ref{vel}) with the initial position $r_0=\sigma_0$ we get:
\begin{equation}\label{time}
t=\pm \frac{1}{\sqrt{2}\left(\frac{2}{\pi}\right)^{\frac{1}{4}}} \left(\frac{\sigma_0^3}{Gm}\right)^{\frac{1}{2}} \int_{\sigma_0}^{r} \frac{dr^{\prime}}{\left(\sigma_0^2-r^{\prime^2}+\frac{r^{\prime^4}}{2\sigma_0^2} \right)^{\frac{1}{2}}}=\pm \frac{1}{\sqrt{2}\left(\frac{2}{\pi}\right)^{\frac{1}{4}}} \left(\frac{\sigma_0^3}{Gm}\right)^{\frac{1}{2}} F\left(\frac{r}{\sigma_0}\right)
\end{equation}
as the falling time from $\sigma_0$ to the $r$. The function $F\left(\frac{r}{\sigma_0}\right)$ denotes an elliptical function which has finite value. The values of the elliptical integrals are found in the tables of integrals. Because of deterministic structure of Bohmian mechanics, we conclude that our approach is also deterministic. The only problem is that we do not have enough information about the exact value of $r_0$ at $t=0$. In Bohm's own view the position of the particle is a hidden variable. In standard quantum mechanics the definition of trajectory is not possible and all information about the system is contained only in the wave function of the system. Thus, such pictureable approach is not possible there.\par 
We can derive a more rounded satisfactory result. In the relation (\ref{time}), we consider up to second order $r^2$. Hence, we have:
\begin{eqnarray}\label{t}
t &=&\pm \frac{1}{\sqrt{2}\left(\frac{2}{\pi}\right)^{\frac{1}{4}}} \left(\frac{\sigma_0^3}{Gm}\right)^{\frac{1}{2}} \int_{r^{\prime}=\sigma_0}^{r^{\prime}=r} \frac{dr^{\prime}}{\left(\sigma_0^2-r^{\prime^2}\right)^{\frac{1}{2}}}\\ \nonumber 
&=&\pm \frac{1}{\sqrt{2}\left(\frac{2}{\pi}\right)^{\frac{1}{4}}} \left(\frac{\sigma_0^3}{Gm}\right)^{\frac{1}{2}} \arccos \left(\frac{r}{\sigma_0}\right)
\end{eqnarray}
Where, we have used the change of the variable $r^{\prime}=\sigma_0 \cos \theta$ within the integral. Thus, we have a finite value multiplied by $\left(\frac{\sigma_0^3}{Gm}\right)^{\frac{1}{2}}$ again. 
Now, we have from the relation (\ref{t}) that:
\begin{equation}\label{cos}
r(t)=\sigma_0 \cos \left[\sqrt{2}\left(\frac{2}{\pi}\right)^{\frac{1}{4}}\left(\frac{Gm}{\sigma_0^3}\right)^{\frac{1}{2}}t \right], \quad t \approx 0
\end{equation}
Note that the relation (\ref{cos}) does not imply on a periodic behavior for all $t$. Because, this relation has been derived for approximate $t\approx0$ or $\frac{\hbar^2 t^2}{4m^2 \sigma_0^4} \ll 1$. Now, the relation (\ref{cos}) can be written in the form
\begin{equation}\label{tb}
r(t)=\sigma_0 -\frac{1}{2} \frac{Gm}{\sigma_0^2}t^2 \quad \text{for}\quad  t \approx 0
\end{equation}
where, we have used the expansion ($\cos \left[\left(\frac{Gm}{\sigma_0^3}\right)^{\frac{1}{2}}t \right]=1-\frac{Gm}{2\sigma_0^3}t^2 +\cdots$). Also, we have dropped the ineffective factor $\sqrt{2}\left(\frac{2}{\pi}\right)^{\frac{1}{4}}$ to have a more rounded result.
Here, we can introduce the self-gravitational acceleration or the  distribution acceleration in the form:
\begin{equation}
g_{ \psi }=\frac{Gm}{\sigma_0^2}
\end{equation}
The reduction time $\tau$, is obtained by substituting the $r(t=\tau)=0$, into the relation (\ref{tb}). Thus we  have:
\begin{equation}\label{time1}
\tau= \left(\frac{\sigma_0^3}{Gm}\right)^{\frac{1}{2}}
\end{equation}
 Since, we have considered that $\sigma(t) \approx \sigma_0$ during the reduction, we can substitute 
 $\sigma_0=\frac{\hbar^2}{Gm^3}$ into the (\ref{time1}) to get
\begin{equation}\label{time2}
\tau=\frac{\hbar^3}{G^2m^5}
\end{equation}
as a second relation for the reduction time.
When $m\rightarrow \infty$, we have $\sigma_0 \rightarrow 0$, then $\rho \rightarrow \delta(r)$ and $\tau \rightarrow 0$. This is what happens for macro objects of our classical localized world. 
\textit{Thus, for the objects we see in the world around us, the reduction time is zero practically. In other words, we always see the object in the center of its wave packet and gravity does not allow the fluctuations in the position of the particle or object. By decreasing the mass, such fluctuations will be appeared by the quantum force}.\par
Now, we compare this reduction time with the results of previous works of some authors.
By using the criterion $\sigma_0=\frac{\hbar^2}{Gm^3}$, it is easy to check that the relations (\ref{time1}) and (\ref{time2}) are equal to that of Karolyhazy and Frenkel i.e, $\tau =\frac{m\sigma_0^2}{\hbar}$ for a point particle.\cite{RefBassi,RefFl}. \par 
Also, in the works of Diosi and Penrose, the reduction time is estimated by the relation $\tau =\frac{\hbar}{\Delta E}=\frac{\hbar}{\Delta U}$, where, $\Delta E$ or $\Delta U$ denotes the uncertainty of self-gravitational energy of the object at two different points.\cite{RefD2,RefP1}. In the works of Diosi and Penrose the reduction time refers to the decay time in which, two different states reduce to a single state. A quick examination by using the equation (\ref{pot2}), demonstrates that
\begin{equation}
\tau = \frac{\hbar}{\vert U_{g}^{(s)}(\sigma_0)-U_{g}^{(s)}(0)\vert}\approx \frac{\hbar^3}{G^2m^5}= \left(\frac{\sigma_0^3}{Gm}\right)^{\frac{1}{2}}
\end{equation}
This investigations convince us about the  accuracy and correctness of the relations (\ref{time1}) and (\ref{time2}). 
It is clear that when $m\rightarrow 0$ (theoretically), then $\tau \rightarrow \infty$. In this situation, it takes infinite time for the particle to get the center of the wave packet. This is what we expect because lighter particle will be affected by the repulsive quantum force further and they have a wider distribution than heavier particles. 
In, the ref \cite{RefBassi} for a proton ($m_p=1.67 \times 10^{-27} kg$), the value of the characteristic width and the reduction time are $\sigma_0 \approx 10^{25} cm $ and $\tau = \frac{m\sigma_0^2}{\hbar}\approx 10^{53} s$. The relations (\ref{time1}) and (\ref{time2})also give these results. 
\par
It is also possible to define an average reduction time in the ensemble of trajectories as
\begin{equation}
\langle\tau \rangle=\frac{\sigma_0}{\sqrt{\langle u^2 \rangle}}=\frac{\sigma_0}{\sqrt{\int_{0}^{\infty} \rho u^2 4\pi r^2 dr}} \approx \left(\frac{\sigma_0^3}{Gm}\right)^{\frac{1}{2}}
\end{equation}
where we have substituted the relation (\ref{vvv}) into the velocity $u$ under the radical. 
This new interpretation about the wave function reduction is possible in the framework of Bohm's  quantum theory and can be studied for further developments. The Bohmian analysis leads to the equation of motion i.e, (\ref{bb1}) which gives us exact numeric results for the reduction time. Also, it provides a clear picture of the reduction process. We have already examined the role of gravity in the wave function reduction in the Bohmian context in refs \cite{RefRGG1} and \cite{RefRGG2}. But, in this study we investigated it in more detail and we obtained further new results. 
\section{conclusion}
\label{sec:3}
By using the concepts of Bohmian quantum mechanics, we provided a dynamical and deterministic framework to study the problem of gravity induced wave function reduction. This was done for a point particle and it can be developed for extend objects. We argued that the self-gravity  which is defined on the basis of quantum distribution can prevents more dispersion or divergence of trajectories. The threshold of classical world is determined by the equality of the self-gravitational and quantum forces from which we derived the famous criterion $\sigma_0=\frac{\hbar^2}{Gm^3}$. We studied the dynamics of the particle during the wave function reduction (gravity-dominant regime) and we derived the interesting equation of motion (\ref{bb1}). The numerical results of reduction equation was investigated. Then, an approximate  satisfactory analytic solution which gives a formula for the reduction time was derived. Since, Bohmian mechanics is a deterministic theory, we conclude that the reduction process or breaking the quantum superposition which means the dominance of gravitational force over the quantum force is  also a clear dynamical processes and suddenly jump from the quantum world to the classical world can be justified by a deterministic approach. But, due to our ignorance about the initial data, we can not predict the evolution of the system exactly and we have to use ensemble description. When, the particle arrives the center of wave packet we do not know it was on which of the trajectories. The reduction time is defined as the time needed for a particle to get the center of the wave packet or the Gaussian distribution under the effects of gravity. Instead of a sharp distinction between the classical and quantum worlds we can assert that for the classical world we have $f_g \gg f_q$. Therefore, the uncertainty in the position of the particle becomes negligible and its wave function can be considered to be a Dirac delta function practically. 
This interpretation for the reduction process and the reduction time is a novel work and can help us to develop the study of wave function reduction in a deterministic framework. Also, a generalization to the relativistic and geometric concepts in which gravity can be considered as space time effects, may provide a clue to find more fundamental concepts. 
\section*{Acknowledgement}
We thank Dr Ghadir Jafari for his useful comments and helpful discussions.


\begin{thebibliography}{}
%
%
\bibitem{RefK1}
Károlyházy.F, Gravitation and quantum mechanics of macroscopic objects, Nuovo Cimento A (1965-1970) (1966) 42: 390.
\bibitem{RefK2}
Károlyházy, F.: Gravitation and quantum mechanics of macroscopic bodies. Magyar Fizikai Polyoirat
12, 24 (1974)
\bibitem{RefK3}
Károlyházy, F., Frenkel, A., Lukács, B.: On the possible role of gravity on the reduction of the wave
function. In: Penrose, R., Isham, C.J. (eds.) Quantum Concepts in Space and Time, pp. 109–128.
Oxford University Press, Oxford (1986)
\bibitem{RefD1}
Diósi.L, Gravitation and quantum-mechanical localization of macro-objects, Physics Letters A,Volume 105, Issues 4–5 (1984).
\bibitem{RefD2}
Diósi.L, A universal master equation for the gravitational violation of quantum mechanics, Physics Letters A, Volume 120, Issue 8, (1987).
\bibitem{RefD3}
Diósi, L, Models for universal reduction of macroscopic quantum fluctuations, Phys. Rev. A 40, 1165,
(1989).
\bibitem{RefP1}
Penrose, Roger, On Gravity's Role in Quantum State Reduction. General Relativity and Gravitation, Vol. 28,No.5, 1996
\bibitem{RefP2}
Penrose, Roger,2009 J.Phys.: Conf.Ser.174012001
\bibitem{RefP3}
Penrose, Roger, On the Gravitization of Quantum Mechanics 1: Quantum State Reduction. Found Phys 44:557-575, (2014).
\bibitem{RefB}
Bohm, D., A suggested interpretation of quantum theory in terms of hidden variables I and II. Phys.Rev. 85(2) (1952) 166-193
\bibitem{RefI}
Bohm, D, Wholeness and The Implicate Order, Routledge \& Kegan Paul, (1980)
\bibitem{RefU}
Bohm, D and Hiley,B.J, The undivided universe:An ontological interpretation of quantum theory. Routledge, (1993)
\bibitem{RefH}
Holland P. R. ,The Quantum Theory of Motion. Cambridge, Cambridge University Press (1993).
\bibitem{RefC}
Cushing J.T. The Causal Quantum Theory Program. In: Cushing J.T., Fine A., Goldstein S. (eds) Bohmian Mechanics and Quantum Theory: An Appraisal. Boston Studies in the Philosophy of Science, vol 184.(1996).
\bibitem{RefPet}
Peter J. Riggs, Quantum Causality: Conceptual Issues In The Causal Theory Of Quantum Mechanics. Springer, Dordrecht Heidelberg London New York (2009).
\bibitem{RefM}
 Mario Bunge, Causality; the Place of the Causal Principle in Modern Science,  Harvard University; Ex-seminary Library edition (1959).
 \bibitem{RefBal}
Belinfante, F.J, A Survey of Hidden-variables Theories, Elsevier Science and Technology, (1973).
\bibitem{RefBell}
Bell. J. B. (1966) “On the problem of hidden variables in quantum mechanics,” Reviews of Modern Physics 38 (3): 447
\bibitem{RefGir1}
G. C. Ghirardi, A. Rimini, T. Weber. A model for a unified quantum description of macroscopic and microscopic systems, Quantum Probability and Applications II pp 223-232, (1984).
\bibitem{RefGir2}
Ghirardi, G.C., Rimini, A., and Weber, T. Unified dynamics for microscopic and macroscopic systems, Phys. Rev. D 34, 470 .( 1986).
\bibitem{RefBassi}
A. Bassi, K. Lochan, S. Satin, T. P. Singh, and H. Ulbricht, Models of wave-function collapse, underlying theories, and experimental tests. Rev. Mod. Phys. 85, 471. (2013).
\bibitem{RefShoja}
Shojai. F, Golshani, M.  On The Geometrization Of Bohmian Mechanics: A New Approach To Quantum Gravity. International Journal of Modern Physics A, Vol. 13, No. 4 (1998).
\bibitem{RefAtigh}
Atiq, M., Karamian, M. Golshani, M. A New Way for the Extension of Quantum Theory: Non-Bohmian Quantum Potentials. Found Phys 39, 33–44 (2009).
\bibitem{RefMach}
Rahmani, F., Golshani, M. Some Clarifications on the Relation Between Bohmian Quantum Potential and Mach’s Principle. Int J Theor Phys 56, 3096–3107 (2017).
\bibitem{RefFl}
Frenkel, A. A Tentative Expression of the Károlyházy Uncertainty of the Space-Time Structure Through Vacuum Spreads in Quantum Gravity. Foundations of Physics 32, 751–771 (2002).
\bibitem{RefRGG1}
F.Rahmani, M. Golshani, Gh. Jafari, Gravitational reduction of the wave function based on Bohmian quantum potential. International Journal of Modern Physics A, Volume 33, Issue 22. (2018).
\bibitem{RefRGG2}
F.Rahmani, M. Golshani, Gh. Jafari, A geometric look at the objective gravitational wave function reduction. Pramana-J Phys 94,163(2020).
\end{thebibliography}
\end{document}